\documentclass[aps,twocolumn,pra,superscriptaddress,amsmath,showpacs,tightenlines]{revtex4}
\usepackage{epsfig,graphicx,times}
\usepackage{amstext}
\usepackage{amsmath}            
\usepackage{amssymb}            
\usepackage{graphicx}           
\usepackage{latexsym}
\usepackage{bm}
\usepackage[colorlinks,citecolor=blue, linkcolor=blue,hyperindex,CJKbookmarks,dvipdfm]{hyperref}

\begin{document}

\title{Mechanical $\mathcal{PT}$ symmetry in coupled optomechanical systems}
\author{Xun-Wei Xu}
\affiliation{Beijing Computational Science Research Center, Beijing 100094, China}
\author{Yu-xi Liu}
\affiliation{Institute of Microelectronics, Tsinghua University, Beijing 100084, China}
\affiliation{Tsinghua National Laboratory for Information Science and Technology
(TNList), Beijing 100084, China}
\author{Chang-Pu Sun}
\affiliation{Beijing Computational Science Research Center, Beijing 100094, China}
\affiliation{Synergetic Innovation Center of Quantum Information and Quantum Physics,
University of Science and Technology of China, Hefei, Anhui 230026, China}
\author{Yong Li}
\email{liyong@csrc.ac.cn}
\affiliation{Beijing Computational Science Research Center, Beijing 100094, China}
\affiliation{Synergetic Innovation Center of Quantum Information and Quantum Physics,
University of Science and Technology of China, Hefei, Anhui 230026, China}
\date{\today}

\begin{abstract}
We propose to realize mechanical parity-time ($\mathcal{PT}$) symmetry in two
coupled optomechanical systems. To provide gain to one mechanical resonator
and the same amount of damping to the other, the two optical cavities should
be driven by blue- and red-detuned laser fields, respectively. After
adiabatically eliminating the degrees of freedom of the cavity modes, we derive a formula to describe the $\mathcal{PT}$ symmetry of two coupled mechanical resonators. Mechanical $\mathcal{PT}$-symmetric phase transition is demonstrated by the dynamical behavior of the mechanical resonators.
Moreover, we study the effect of the quantum noises on the dynamical behavior of the mechanical resonators when the system is in the quantum regime.
\end{abstract}

\maketitle



\section{Introduction}

In quantum mechanics, the Hamiltonian of a closed system is usually required to be Hermitian, which guarantees real energy spectrum and thus unitary time evolution. Recently, it was found that the axiom of Hermiticity in quantum mechanics could be replaced by the condition of parity-time ($\mathcal{PT}$) symmetry and then complex quantum mechanics was builded~\cite{BenderPRL98,BenderPRL02}.
As the time operator is anti-linear, the eigenstates of the Hamiltonian may or may not be eigenstates of $\mathcal{PT}$ operator, despite the fact that they commute with each other~\cite{KottosNP10}. It has been demonstrated that a threshold exists in the system. Below the threshold, the Hamiltonian has completely real eigenvalues and shares the same set of eigenvectors with the $\mathcal{PT}$ operator. Above the threshold, the eigenvalues are no longer completely real and instead become complex, and the eigenfunctions of them are different from each other. This threshold marks the boundary between the unbroken and broken $\mathcal{PT}$ symmetries. In the complex quantum mechanics, if a system has an unbroken $\mathcal{PT}$ symmetry, then it will have positive probabilities and is subject to unitary time evolution by constructing a new type of inner product~\cite{BenderPRL02}. Non-Hermitian $\mathcal{PT}$-symmetric Hamiltonians play a significant role in complex quantum mechanics and quantum field theory (for a review, see Ref.~\cite{BenderRPP07}).

Although the non-Hermitian based complex quantum mechanics is still debated, experimentalists are trying to test the $\mathcal{PT}$ symmetry in non-Hermitian systems. In particular, optical systems with complex refractive indices provide an appropriate platform for this study~\cite{El-GanainyOL07,MakrisPRL08,LuoPRL13}.
The $\mathcal{PT}$ symmetry has been experimentally demonstrated in two coupled waveguides~\cite{RuterNP10}, photonic lattices~\cite{SzameitPRA11,RegensburgerNat12}, microwave billiard~\cite{BittnerPRL12} or transmission line~\cite{SunPRL14}, and whispering-gallery microcavities~\cite{BoPengNP14,BenderPRA13,JingPRL14}. The optical systems with $\mathcal{PT}$ symmetry have many important applications, for example, non-reciprocal light propagation~\cite{RuterNP10,FengSci11,BoPengNP14}, double refraction~\cite{MakrisPRL08}, absorption-enhanced transmission~\cite{GuoPRL09}, coherent perfect absorber~\cite{LonghiPRA10,ChongPRL11,SunPRL14}, and unidirectional invisibility~\cite{LinPRL11,RegensburgerNat12}. In addition, the $\mathcal{PT}$ symmetry has also been realized in the active \emph{LRC} circuits~\cite{SchindlerPRA11,LinPRA12}.

It is well known that the system of coupled mechanical oscillators is one of the most direct and simplest systems to illustrate non-Hermitian $\mathcal{PT}$ symmetry. Recently, Bender~\emph{et al.}~\cite{BenderAJP13} experimentally demonstrated the $\mathcal{PT}$ phase transition in a simple classical mechanical system
of two coupled pendulums with controllable damping and gain respectively, where the kinetic
energy are added to or subtracted from the coupled pendulums by an electromagnet with brief impulses. When the the damping and gain parameters of the two pendulums are below a critical value, the system is in the unbroken-$\mathcal{PT}$-symmetric region.

Recently, it was shown in experiments that mechanical resonators can be coupled to the electromagnetic fields via radiation pressure or optical gradient forces in so-called optomechanical systems, which have drawn much attention in the past decade. Optomechanical systems can be used to produce {nonclassical states of mechanical modes}~\cite{XuPRA13a,XuPRA13b,TanPRA13} and they also have the potential applications in quantum information processing (for reviews, see Refs.~\cite{Kippenberg,Marquardt,Aspelmeyer}). It has been theoretically studied and experimentally demonstrated that the motion of a mechanical resonator in the optomechanical system can be controlled by driving the optical cavity with an external laser field. If the frequency of the laser field becomes lower than the cavity's resonant frequency (red-detuned case), the motion of the mechanical resonator can be suppressed and cooled down~\cite{Gigan, Arcizet,Schliesser,Wilson-RaePRL07, MarquardtPRL07, TeufelPRL, Thompson,Schliesser2008,Rocheleau, Groblacher2009, Park, Schliesser2009,
Teufel475,Chan2009, Verhagen,YLiPRA08,jqliao,NiePRA13}. On the contrary, if the laser field is tuned above resonance (blue-detuned case), the motion of the resonator will be enhanced and heated~\cite{BraginskyPLA01,KellsPLA02,BraginskyPLA02,RokhsariOE05,KippenbergPRL05,MarquardtPRL06,MetzgerPRL08,LudwigNJP08,QianPRL12,AnetsbergerNaP09,RodriguesPRL10,ArmourCRP12,NationPRA13,LorchARX13,GrudininPRL10,WuNJP13,KhurginPRL12,KhurginNJP12,PootPRA13}. Thus we can control the damping (gain) of the mechanical resonators by driving the optomechanical cavities with red- (blue-) detuned laser fields. This provides us the most important condition to realize mechanical $\mathcal{PT}$ symmetry by optomechanical systems.

Here, we propose to realize the $\mathcal{PT}$ symmetry by two coupled mechanical resonators by virtue of two optomechanical systems. In contrast to Ref.~\cite{BenderAJP13}, here the gain and damping of the mechanical resonators are controlled by driving the cavities with laser fields, thus the gain-damping ratio can be balanced by adjusting the pump powers. Moreover, we show the dynamical behavior of the mechanical resonators when the system is in the quantum regime and the quantum noises are taking into account by the quantum Langevin equations~\cite{AgarwalPRA13,DastPRA14,HePRA15}. In recent years, the mechanical resonators of the optomechanical systems have been cooled and prepared in the ground states~\cite{Teufel475,Chan2009}, and our proposal may pave the way to study the mechanical $\mathcal{PT}$-symmetric systems in the quantum regime.

The paper is organized as follows: In Sec.~II, the Hamiltonian of the
coupled optomechanical systems is introduced and the $\mathcal{PT}$-symmetric equations for the mechanical modes are derived. The mechanical $\mathcal{PT}$-symmetric phase transition is shown in Sec.~III. In Sec.~IV, we analyze the effect of the quantum noises on the dynamical behavior of the mechanical resonators when the system is in the quantum regime. Finally we draw our conclusions in Sec.~V.

\section{Model and $\mathcal{PT}$-symmetric equations}

\begin{figure}[tbp]
\includegraphics[bb=18 220 573 610, width=8.5 cm, clip]{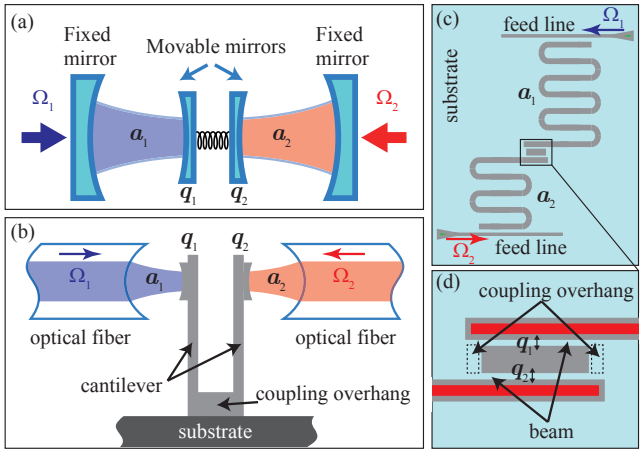}
\caption{(Color online) (a) Schematic diagram of two coupled optomechanical systems
with the cavities being driven by blue- and red-detuned laser fields respectively. (b), (c) and (d) The possible realistic physical systems to implement the setup of coupled optomechanical systems for realizing mechanical parity-time ($\mathcal{PT}$) symmetry: (b) the optical fiber cavity with one-end vibrating cantilever~\cite{ShkarinPRL14}; (c) and (d) the superconducting transmission line resonator coupled to a mechanical beam~\cite{RegalNat08,Rocheleau,MasselNat11}.}
\label{fig1}
\end{figure}

\begin{figure}[tbp]
\includegraphics[bb=76 153 504 716, width=6 cm, clip]{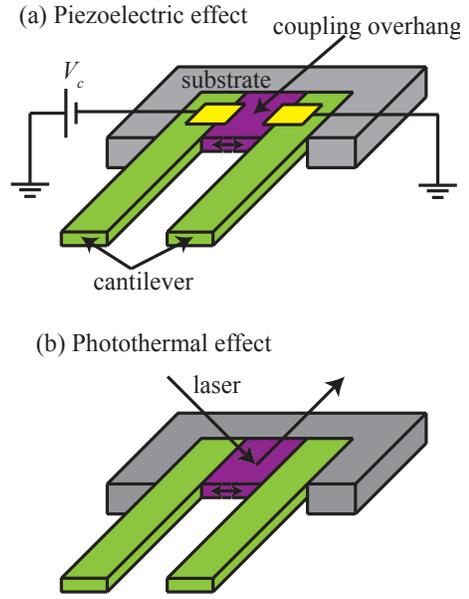}
\caption{(Color online) Tunable coupling of two vibrating cantilever by (a) piezoelectric effect and (b) photothermal effect.}
\label{fig2}
\end{figure}

As schematically shown in Fig.~\ref{fig1}(a), we consider a schematic model consisting of two optomechanical systems with two vibrating mirrors coupling to each other~\cite{KumarOC12}. Experimentally, the optomechanical system required in our proposal can be realized by the optical fiber cavity with one-end vibrating cantilever~\cite{ShkarinPRL14} [Fig.~\ref{fig1}(b)], or the superconducting transmission line resonator coupled to a mechanical beam~\cite{RegalNat08,Rocheleau,MasselNat11,HeinrichEPL11} [Figs.~\ref{fig1}(c) and (d)].

In Fig.~\ref{fig1}(b), the optical fiber cavity is formed between the end face of a single-mode optical fiber and a vibrating cantilever; the fiber face has a concavity with a highly reflective dielectric coating~\cite{ShkarinPRL14} and the cantilever is coated with a high-reflectivity Bragg mirror~\cite{Groblacher2009}. The mechanical coupling between the cantilevers can be obtained via a common base with a separating distance corresponding to the coupling overhang~\cite{SpletzerAPL06,Gil-SantosAPL11}.

The coupled optomechanical systems can also be realized in the microwave domain by embedding a nanomechanical beam inside a superconducting transmission line microwave cavity~\cite{RegalNat08,Rocheleau,MasselNat11,HeinrichEPL11}, as schematically shown in Fig.~\ref{fig1}(c). The area framed by the dashed line in Fig.~\ref{fig1}(d) shows the coupling overhang, which provides the mechanical coupling between the two beams~\cite{KarabalinPRB09,OkamotoAPE09,OkamotoPE10,OkamotoAPL11,OkamotoNaP13}. Different from the system given in Ref.~\cite{HeinrichEPL11} where the coupling between the superconducting microwave resonators plays a important role, here such a coupling (cross-talk) should be avoided in the system under consideration, e.g., by employing two microwave cavities far off-resonant with each other or well separated in space.

Experimentally, the mechanical coupling between the two cantilevers can be controlled by applying stress to the coupling overhang by piezoelectric effect or photothermal effect~\cite{KarabalinPRB09,OkamotoAPE09,OkamotoPE10,OkamotoAPL11,OkamotoNaP13}.
As shown in Fig.~\ref{fig2}(a), applying a dc voltage to the electrodes [yellow areas in Fig.~\ref{fig2}(a)], the effective spring constant of the coupling overhang is changed via the stress generated by the piezoelectric effect and therefore the mechanical coupling constant is varied. Another alternative way for mechanical coupling modulation is given by the photothermal stress, which is induced by the irradiation of the laser [Fig.~\ref{fig2}(b)].
The optically induced thermal stress or the tension coupling overhang is proportional to the laser power. By adjusting the laser power, one can effectively control the mechanical coupling between the two cantilevers. In addition, the two vibrating mirrors can also be coupled through the electrostatic force~\cite{HuangPRL13} or Coulomb interaction for charged vibrating mirrors~\cite{HensingerPRA05,ZhangPRA12,RenarX14}. For the sake of generality, in the following study, we will not specify our theoretical model to any particular system.

The Hamiltonian of the coupled optomechanical systems can be given by ($\hbar=1$)
\begin{equation}
H=H_{om}+H_{c}+H_{d},
\end{equation}%
where
\begin{eqnarray}
H_{om} &=&\sum_{i=1,2}\left[ \Delta _{i}a_{i}^{\dag }a_{i}+\frac{\omega
_{i}}{2}\left( q_{i}^{2}+p_{i}^{2}\right) -g_{i}a_{i}^{\dag }a_{i}q_{i}\right] ,
\\
H_{c} &=&-Jq_{1}q_{2}, \\
H_{d} &=&\sum_{i=1,2}\left( \Omega _{i}a_{i}^{\dag }+\mathrm{H.c.}\right) .
\end{eqnarray}%
$H_{om}$ describes the Hamiltonian of two standard optomechanical systems; $a_{i}$, and $%
a_{i}^{\dag }$ ($i=1,2$) are the annihilation and creation operators of the
cavity mode with frequency $\omega _{i}^{c}$; the vibrating mirrors ($q_{i}$
and $p_{i}$, the dimensionless position and momentum operators for the
vibrating mirrors) act as two mechanical resonators with mechanical frequency $%
\omega _{i}$; $g_{i}$ is the single-photon optomechanical coupling strength between the $i$th cavity mode and $i$th mechanical resonator. $H_{c}$ describes the coupling between the two mechanical resonators with the coupling strength $J$ which is assumed to be much smaller than the mechanical frequency ($\omega _{i}\gg J$). As given in the term $H_{d}$, the two single-mode cavities are driven by two external fields with the driving strengths $\Omega _{i}$, respectively; $\Delta _{i}=\omega _{i}^{c}-\omega_{i}^{d}$ is the frequency detuning between the $i$th cavity mode ($\omega _{i}^{c}$) and $i$th external driving field ($\omega_{i}^{d}$).

The dynamics of the cavity fields and mechanical oscillators can be
described by the quantum Langevin equations. After considering the dissipations
but neglecting the fluctuations of the cavity fields and mechanical
resonators in the strong external driving condition, we can obtain the equations
\begin{eqnarray}
\frac{d}{dt}a_{i}&=&-\left[ \frac{\kappa _{i}}{2}+i\left(
\Delta_{i}-g_{i}q_{i}\right) \right] a_{i}-i\Omega _{i},  \label{eq:5} \\
\frac{d}{dt}q_{i}&=&\omega _{i}p_{i}, \\
\frac{d}{dt}p_{1}&=&-\omega _{1}q_{1}+Jq_{2}+g_{1}a_{1}^{\dag }a_{1}-\frac{%
\gamma _{1}}{2}p_{1}, \\
\frac{d}{dt}p_{2}&=&-\omega _{2}q_{2}+Jq_{1}+g_{2}a_{2}^{\dag }a_{2}-\frac{%
\gamma _{2}}{2}p_{2},  \label{eq:10}
\end{eqnarray}
for $i=1,2$. Here $\kappa _{i}$ is the decay rate of the $i$th cavity and
and $\gamma _{i}$ is the damping rate of the $i$th mechanical resonator. To
solve the above nonlinear dynamical equations, we can write each operator as
the sum of its steady-state value and the time-dependent term: $a_{i}
\rightarrow \alpha _{i}+a_{i} $ and $q_{i} \rightarrow \xi_{i}+q_{i}$, where
$\alpha _{i}$ and $\xi_{i}$ are the steady-state values of the system and
satisfy the following equations:
\begin{eqnarray}
\left[ \frac{\kappa _{i}}{2}+i\left( \Delta _{i}-g_{i}\xi_{i}\right) \right]
\alpha _{i} &=& -i\Omega _{i},  \label{eq:11} \\
\omega _{1}\xi_{1}-J\xi_{2} &=& g_{1}\left\vert \alpha _{1}\right\vert ^{2},
\\
\omega _{2}\xi_{2}-J\xi_{1} &=& g_{2}\left\vert \alpha _{2}\right\vert ^{2}.
\label{eq:14}
\end{eqnarray}
In the strong external driving condition $\Omega _{i} \gg \kappa _{i}$, one
has $|\alpha _{i}|^2 \gg 1$. Thus the nonlinear terms in Eq.~(\ref{eq:5})-(\ref{eq:10}) (e.g. $g_{i}a_{i}^{\dag }a_{i}$) can be neglected and the linearized equations for time-dependent terms are given as
\begin{eqnarray}
\frac{d}{dt}a_{i}&=&-\left( \frac{\kappa _{i}}{2}+i\Delta_{i}^{\prime
}\right) a_{i}+iG _{i}q_{i},  \label{eq:15} \\
\frac{d}{dt}q_{i}&=&\omega _{i}p_{i},  \label{eq:17} \\
\frac{d}{dt}p_{1}&=&-\omega _{1}q_{1}+Jq_{2}+G
_{1}^{\ast}a_{1}+G_{1}a_{1}^{\dag }-\frac{\gamma _{1}}{2}p_{1},
\label{eq:19} \\
\frac{d}{dt}p_{2}&=&-\omega _{2}q_{2}+Jq_{1}+G
_{2}^{\ast}a_{2}+G_{2}a_{2}^{\dag }-\frac{\gamma _{2}}{2}p_{2},
\label{eq:20}
\end{eqnarray}
where $\Delta_{i}^{\prime }=\Delta_{i}-g_{i}\xi_{i}$ is the effective optical
detuning and the parameter $G_{i}=g_{i}\alpha _{i}$ represents the effective
optomechanical coupling constant.

From Eqs.~(\ref{eq:15})-(\ref{eq:20}), we can derive the $\mathcal{PT}$-symmetric dynamical equations for the coupled mechanical resonators. Under the assumption that the decay rates of the cavities are much larger than the effective optomechanical coupling, $\kappa _{i} \gg G _{i}$, we can adiabatically eliminate the cavity modes~\cite{JahnePRA09} (for details see Appendix A), then we find
\begin{eqnarray}
\frac{d}{dt}p_{1} &=&-\left( \omega _{1}+\delta \omega _{1}\right)
q_{1}+Jq_{2}+\frac{1}{2} \left( \Gamma _{1}-\gamma _{1}\right) p_{1},
\label{eq:27} \\
\frac{d}{dt}p_{2} &=&-\left( \omega _{2}-\delta \omega _{2}\right)
q_{2}+Jq_{1}-\frac{1}{2} \left( \Gamma _{2}+\gamma _{2}\right) p_{2},
\label{eq:28}
\end{eqnarray}%
where
\begin{eqnarray}
\delta \omega _{i} &=&\frac{8\left\vert G_{i}\right\vert ^{2}\omega _{i}}{
\kappa _{i}^{2}+16\omega _{i}^{2}}, \label{eq:29} \\
\Gamma _{i} &=&\frac{4\left\vert G_{i}\right\vert ^{2}}{\kappa _{i} }\frac{%
16\omega _{i}^{2}}{\kappa _{i}^{2}+16\omega _{i}^{2}}, \label{eq:30}
\end{eqnarray}
are the radiation pressure induced frequency shift and gain (or damping)~%
\cite{Wilson-RaePRL07,MarquardtPRL07}. In the resolved-sideband regime $%
\omega _{i} \gg \kappa _{i}$ and the adiabatic elimination conditions $%
\kappa _{i} \gg G _{i}$, the frequency shift induced by the radiation
pressure is very small ($\delta \omega _{i} \ll \omega _{i}$). If the
external driving fields are strong enough, then the original mechanical
damping rates will be much smaller than the radiation pressure induced gain
(damping) $\gamma _{i} \ll \Gamma _{i}$. After omitting the negligible
frequency shift $\delta \omega _{i}$ and original mechanical damping $\gamma
_{i}$, and taking the degenerate parameters of mechanical resonators: $%
\gamma _{\mathrm{eff}}=\Gamma_{1}=\Gamma _{2}$ and $\omega _{m}=\omega
_{1}=\omega _{2}$, one can get the dynamical equations for the coupled
mechanical resonators with $\mathcal{PT}$ symmetry~\cite{BenderAJP13}
\begin{eqnarray}
\frac{d}{dt}q_{1} &=& \omega _{m}p_{1},  \label{eq:31} \\
\frac{d}{dt}q_{2} &=& \omega _{m}p_{2},  \label{eq:32} \\
\frac{d}{dt}p_{1} &=& -\omega _{m}q_{1}+Jq_{2}+\frac{\gamma _{\text{eff}}}{2}%
p_{1},  \label{eq:33} \\
\frac{d}{dt}p_{2} &=& -\omega _{m}q_{2}+Jq_{1}-\frac{\gamma _{\text{eff}}}{2}%
p_{2},  \label{eq:34}
\end{eqnarray}
Eqs.~(\ref{eq:31}-\ref{eq:34}) can also be written in an equivalent form as
\begin{eqnarray}
\frac{d^{2}}{dt^{2}}q_{1}-\frac{\gamma _{\text{eff}}}{2}\frac{d}{dt}q_{1}
&=& -\omega _{m}^{2}q_{1}+\omega _{m}Jq_{2},  \label{eq:35a} \\
\frac{d^{2}}{dt^{2}}q_{2}+\frac{\gamma _{\text{eff}}}{2}\frac{d}{dt}q_{2}
&=& -\omega _{m}^{2}q_{2}+\omega _{m}Jq_{1}.  \label{eq:36a}
\end{eqnarray}
It is ready to check that the dynamical equations [Eqs.~(\ref{eq:31})-(\ref%
{eq:34}) or Eqs.~(\ref{eq:35a})-(\ref{eq:36a})] are invariant under the $%
\mathcal{PT}$ transformation (i.e. $\mathcal{P}$: the subscripts $1
\leftrightarrow 2$; $\mathcal{T}$: $t \rightarrow -t$, $p_{i} \rightarrow - p_{i}$).

Now, let us derive the threshold marking the boundary between the broken and
unbroken $\mathcal{PT}$-symmetric regions. Eqs.~(\ref{eq:31})-(\ref{eq:34})
can be rewritten in a compact matrix form as
\begin{equation}
i\frac{d}{dt} \left\vert \Psi \right\rangle =H_{\text{eff}} \left\vert \Psi
\right\rangle,  \label{eq:35}
\end{equation}%
with $\left\vert \Psi \right\rangle =\left( q_{1},p_{1},q_{2},p_{2}\right)
^{T}$, and the effective Hamiltonian
\begin{equation}  \label{eq:36}
H_{\text{eff}}=i\left(
\begin{array}{cccc}
0 & \omega _{m} & 0 & 0 \\
-\omega _{m} & \frac{\gamma _{\mathrm{eff}}}{2} & J & 0 \\
0 & 0 & 0 & \omega _{m} \\
J & 0 & -\omega _{m} & -\frac{\gamma _{\mathrm{eff}}}{2}%
\end{array}
\right).
\end{equation}%
The eigenvalues of the effective Hamiltonian $H_{\mathrm{eff}}$ are given as
\begin{equation}  \label{eq:39}
\lambda_{1,2,3,4} =\pm \lambda_{\pm},
\end{equation}
where
\begin{equation}  \label{eq:40}
\lambda_{\pm} = \sqrt{\omega _{m}^{2}-\frac{\gamma _{\text{eff}}^{2}\pm
\sqrt{ \gamma _{\mathrm{eff}}^{4}-16\omega _{m}^{2}\left( \gamma _{\text{eff}%
}^{2}-4J^{2}\right) }}{8}}.
\end{equation}
As $\omega _{m}\gg J \sim \gamma _{\text{eff}}$, in order to ensure that all the eigenvalues of $H_{\text{eff}}$ are real, the effective damping rate $\gamma _{\text{eff}}$ need to satisfy
\begin{equation}
\gamma _{\text{eff}} ^{4}-16\omega _{m}^{2}\gamma _{\text{eff}}^{2}+64\omega _{m}^{2}J^{2} \geq 0.
\end{equation}%
This condition gives the $\mathcal{PT}$-symmetric region
\begin{equation}
0 \leq \gamma _{\text{eff}} \leq \gamma _{\mathcal{PT}},
\end{equation}
with the $\mathcal{PT}$-symmetric threshold
\begin{equation}
\gamma _{\mathcal{PT}} = 2\sqrt{2}\omega _{m}\sqrt{1-\sqrt{1-\left( J/\omega_{m}\right) ^{2}}} \approx 2J.
\end{equation}
Eq.~(\ref{eq:35}) can be solved analytically by the methods of bi-orthogonal basis~\cite{WongJMP67,SunPS93,LeungPRE98} (for details see Appendix B). In the next section, we are interested in the dynamics of the mean value of the system, thus we treat the operators as c-numbers in the semi-classical approximation.

\section{Mechanical $\mathcal{PT}$-symmetric phase transition}

\begin{figure}[tbp]
\includegraphics[bb=7 306 586 543, width=8.5 cm, clip]{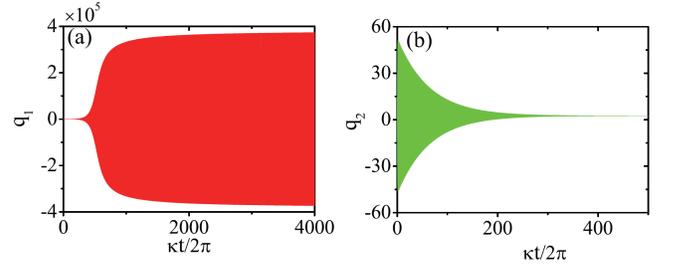}
\caption{(Color online) The dynamical behaviour of the two mirrors for $J=0$:
(a) left mirror $q_{1}$ (red curve) and (b) right mirror $q_{2}$ (green
curve). The parameters used in the numerical calculation are $\omega_{m}=10 \kappa$, $-\Delta_{1}=\Delta_{2}=\omega _{1}=\omega _{2}=\omega _{m}$, $\gamma = \kappa/10^{5}$, $\Omega=5000\kappa$ and $g= \kappa/10000$.}
\label{fig3}
\end{figure}

\begin{figure}[tbp]
\includegraphics[bb=19 8 375 278, width=6 cm, clip]{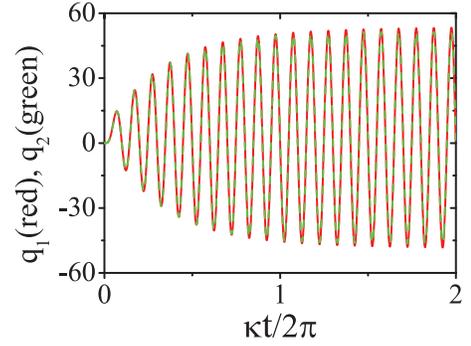}
\caption{(Color online) The dynamical behaviour of the two vibrating mirrors $q_{1}$ (red curves) and $q_{2}$ (green curves) given by Eqs.~(\ref{eq:5})-(\ref{eq:10}) for $J=\kappa/100$. The other parameters used in the numerical calculation are the same as given in Fig.~\ref{fig3}.}
\label{fig4}
\end{figure}

To investigate the transition of $\mathcal{PT}$ symmetry and broken $\mathcal{PT}$ symmetry, we will show the dynamical behaviour of the two mechanical resonators in this section as in Ref.~\cite{BenderAJP13}. Without loss of generality, we assume that $g_{1}=g_{2}=g$, $\kappa_{1}=\kappa_{2}=\kappa$, $\gamma_{1}=\gamma_{2}=\gamma$, $\Omega_{1}=\Omega_{2}=\Omega$ and
normalize all the parameters to $\kappa$. The parameters in the following
numerical calculations are: $\omega_{m}=10 \kappa$, $g=\kappa/10000$, $J=\kappa/100$, $\Delta^{\prime }_{1}=-\omega _{m}$ and $\Delta^{\prime}_{2}=\omega _{m}$. By solving the Eqs.~(\ref{eq:11})-(\ref{eq:14}), we find that: when $\Omega=5000\kappa$, we have $\Delta^{\prime }_{i} \approx \Delta _{i}$, $G _{i} \approx \kappa /20$%
, $\Gamma_{i} \approx \kappa/100$, $\delta \omega _{i} \approx \kappa/8000$.
If the quality factor of the mechanical resonators is high (e.g. $Q_{m}\sim 10^{6}$) so that $\gamma = \kappa/10^{5} \ll \Gamma _{i}$, then we can ignore the original mechanical dampings in the time scale that $t \ll 1/\gamma$.

First of all, we give the dynamical behaviour of the two mirrors in Fig.~\ref{fig3} by solving Eqs.~(\ref{eq:5})-(\ref{eq:10}) numerically in the case that the two mirrors are uncoupled to each other ($J=0$) with initial conditions $q_{1}=p_{1}=q_{2}=p_{2}=0$. In Fig.~\ref{fig3}(b), as the right
cavity (noted by cavity $2$) is driven by a laser resonant to the red sideband, the oscillation amplitude of the right mirror decreases exponentially with rate $\gamma _{\text{eff}}\approx  \kappa/100$. On the contrary, the left cavity (noted by cavity $1$) is driven by a laser resonant to the blue sideband, then the oscillation amplitude of the left mirror increases exponentially with rate $\gamma _{\text{eff}}\approx \kappa/100$ until the  saturation is achieved as shown in Fig.~\ref{fig3}(a). It is the nonlinear terms in the dynamical equations [Eqs.~(\ref{eq:5})-(\ref{eq:10})] that cause the saturation behaviors, and these terms become important as the oscillation amplitude of the mirror increases. The saturation behaviours have already been predicted theoretically~\cite{MarquardtPRL06,LudwigNJP08,QianPRL12,RodriguesPRL10,ArmourCRP12,NationPRA13,LorchARX13,WuNJP13,PootPRA13}
and observed experimentally~\cite{RokhsariOE05,KippenbergPRL05,MetzgerPRL08,AnetsbergerNaP09,GrudininPRL10,KhurginPRL12,KhurginNJP12}.

Due to the adiabatic approximation we have used in the derivation, it is expected that the mechanical $\mathcal{PT}$-symmetric and broken $\mathcal{PT}$-symmetric phases can be observed during the time interval $1/\kappa \ll t \ll 1/\gamma$. Before the time $t$ arrives in this region, there is a transient process for the evolving behavior of the mechanical resonators, as shown in Fig.~\ref{fig4}, which is given by numerically solving Eqs.~(\ref{eq:5})-(\ref{eq:10}) with the initial conditions $q_{1}=p_{1}=q_{2}=p_{2}=0$. It is clear that the time of duration for the transient process is about the lift time of the cavity (e.g. $t=2\pi /\kappa$), and the oscillation amplitudes of the mechanical resonators are about $50$ for $\Omega=5000\kappa$.

\begin{widetext}
\begin{figure*}[tbp]
\includegraphics[bb=11 252 580 663, width=13 cm, clip]{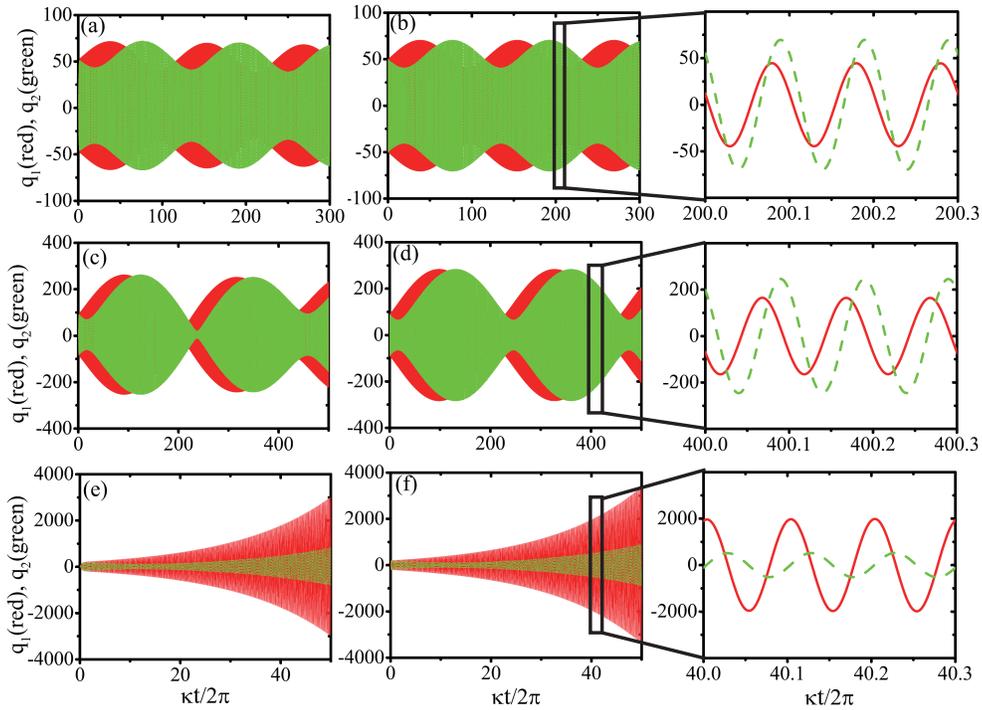}
\caption{(Color online) The dynamical behaviour of the two vibrating mirrors $q_{1}$ (red curves) and $q_{2}$ (green curves) given (a), (c), (e) by Eqs.~(\ref{eq:5})-(\ref{eq:10}) and (b), (d), (f) by Eqs.~(\ref{eq:31})-(\ref{eq:34}) for: (a) $\Omega/\kappa=5000$; (b) $\gamma_{\rm eff} = J$; (c) $\Omega/\kappa=6700$; (d) $\gamma_{\rm eff} = 1.8J$; (e) $\Omega/\kappa=10000$; (f) $\gamma_{\rm eff} = 4J$. The other parameters are the same as that given in Fig.~\ref{fig4}.}
\label{fig5}
\end{figure*}
\end{widetext}

The dynamical behaviour of the two vibrating mirrors for the parameters in the $\mathcal{PT}$-symmetric and broken $\mathcal{PT}$-symmetric region are shown in Fig.~\ref{fig5}(a), (c) and (e) by solving the dynamical equations of the coupled optomechanical system [Eqs.~(\ref{eq:5})-(\ref{eq:10})] directly with driving strength (a) $\Omega/\kappa=5000$ (c) $\Omega/\kappa=6700$ and (e) $\Omega/\kappa=10000$. The corresponding effective damping or gain rates are (a) $\gamma_{\rm eff} \approx J$, (c) $\gamma_{\rm eff} \approx 1.8J$ and (e) $\gamma_{\rm eff} \approx 4J$ according to the Eq.~(\ref{eq:30}).
As comparison, the dynamical behaviours of the two mirrors given by Eqs.~(\ref{eq:31})-(\ref{eq:34}) are shown in Fig.~\ref{fig5}(b), (d) and (f) with effective damping or gain rate (b) $\gamma_{\rm eff} = J$, (d) $\gamma_{\rm eff} = 1.8J$ and (f) $\gamma_{\rm eff} = 4J$.
In order to make the comparison of the results given by Eqs.~(\ref{eq:5})-(\ref{eq:10}) and by Eqs.~(\ref{eq:31})-(\ref{eq:34}) more convenient, we will solve Eqs.~(\ref{eq:5})-(\ref{eq:10}) with the initial conditions $q_{1}=p_{1}=q_{2}=p_{2}=0$ and solve Eqs.~(\ref{eq:31})-(\ref{eq:34}) with the initial conditions $q_{1}=q_{2}=50$ and $p_{1}=p_{2}=0$ in Fig.~\ref{fig5}(b), $q_{1}=q_{2}=90$ and $p_{1}=p_{2}=0$ in Fig.~\ref{fig5}(d), $q_{1}=q_{2}=200$ and $p_{1}=p_{2}=0$ in Fig.~\ref{fig5}(f), respectively.

In Fig.~\ref{fig5}(a) and (b), as $\gamma _{\text{eff}}<2J$, the system is in the $\mathcal{PT}$-symmetric region. The two mirrors become
two beat frequency oscillators with the beat frequency related to the
differences of $\lambda _{+}$ and $\lambda _{-}$ as shown in Eq.~(\ref{eq:40}%
), and they are a little out of phase with each other. In Fig.~\ref{fig5}(c)
and (d), as $\gamma _{\text{eff}}=1.8J$, the system is near the critical point for phase transition (still in the $\mathcal{PT}$-symmetric region),
the oscillation amplitudes increase but the beat frequency becomes lower. In
Fig.~\ref{fig5}(e) and (f), as $\gamma _{\text{eff}}>2J$, the system is in the
broken $\mathcal{PT}$-symmetric region. The oscillation amplitude of the
left mirror increases exponentially, and the oscillation amplitude of the
right mirror also increases after an initial decrease. This is because the
energy in the left mirror is transferred into the right one~\cite{BenderAJP13}.

\begin{figure}[tbp]
\includegraphics[bb=23 380 570 580, width=8.5 cm, clip]{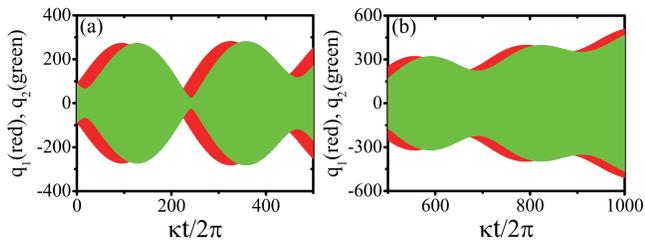}
\caption{(Color online) The dynamical behaviour of the two vibrating mirrors $%
q_{1}$ (red curve) and $q_{2}$ (green curve) given by Eqs.~(\protect\ref%
{eq:17}), (\protect\ref{eq:27}) and (\protect\ref{eq:28}) with $J=\kappa/100$, $\Gamma_{1}-\gamma_{1} = \Gamma _{2}+\gamma _{2} = \gamma _{\rm{eff}}=1.8J$, $\omega _{1}=\omega _{2}=\omega _{m}$ and $\delta \omega _{i}=2.25\times 10^{-5}\kappa$.}
\label{fig6}
\end{figure}

As time goes on, the difference between Fig.~\ref{fig5}(a), (c), (e) and Fig.~\ref{fig5}(b), (d), (f) become more and more significant.
These differences mainly come from the small frequency shift induced by the radiation pressure, $\delta \omega _{i}$.
For the parameters used in our numerical calculation, the frequency shift $\delta \omega _{i}$ is about
$2.25\times 10^{-5}\kappa$ for $\Omega/\kappa=6700$ according to Eq.~(\ref{eq:29}). In Fig.~\ref{fig6}, we show the dynamical behaviour of the two
vibrating mirrors by numerically solving Eqs.~(\ref{eq:17}), (\ref{eq:27}) and
(\ref{eq:28}) and setting $J=\kappa/100$, $\Gamma_{1}-\gamma_{1} \approx \Gamma _{2}+\gamma _{2} \approx \gamma _{\rm{eff}}=1.8J$, $\omega _{1}=\omega _{2}=\omega _{m}$ and $\delta \omega _{i}=2.25\times 10^{-5}\kappa$.
The result [Fig.~\ref{fig6}(a)] agrees well with the result given by Eqs.~(\ref{eq:5})-(\ref%
{eq:10}) as shown in Fig.~\ref{fig5}(c). What is more, this small frequency shift will destroy the $\mathcal{PT}$
symmetry of the system after an enough long time as shown in Fig.~\ref{fig6}(b).
This is different from most of the previous studies that the main difficulty to observe the $\mathcal{PT}$ symmetry was the balance of the gain-damping ratio. As for the ordinary optical systems, the gain and damping of the cavities are difficult to change simultaneously. By contrast, however, in the coupled optomechanical systems, the gain-damping ratio can be easily balanced by adjusting the pump powers in the same time.

\begin{figure}[tbp]
\includegraphics[bb=8 277 583 666, width=8.5 cm, clip]{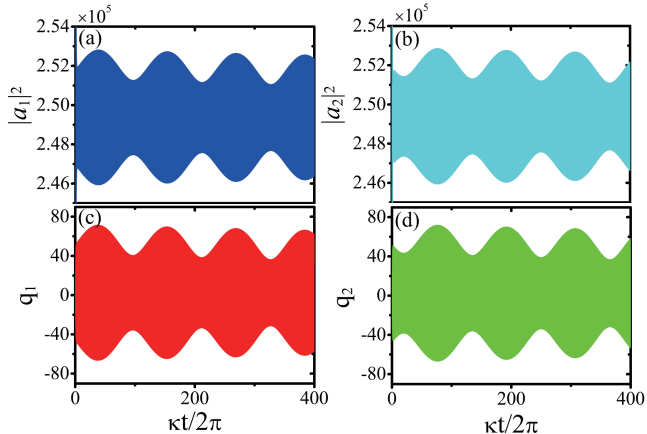}
\caption{(Color online) The dynamical behaviours of the intensity of the optical powers in the cavities and the positions of the two vibrating mirrors given by Eqs.~(\ref{eq:5})-(\ref{eq:10}): (a) $|a_{1}|^{2}$ (blue curve); (b) $|a_{2}|^{2}$ (cyan curve); (c) $q_{1}$ (red curve); (d) $q_{2}$ (green curve). The parameters are the same as that given in Fig.~\ref{fig5}(a).}
\label{fig7}
\end{figure}

For the direct measurement of the dynamical behaviors of the mechanical resonators might not be easy, we propose to demonstrate the mechanical $\mathcal{PT}$ symmetry by measuring the evolution of the output intensity of the cavities. As the decay rates of the cavity fields are much larger than the optomechanical coupling, the intensity of the cavity fields evolve adiabatically with the vibrating mirrors, as shown in Fig.~\ref{fig7}. So the temporal behaviours of the coupled mirrors can be observed by measuring the intensity of the output cavity fields. In the experiment given in Ref.~\cite{CarmonPRL05}, the radiation-pressure-induced mechanical oscillations have been demonstrated by measuring the output intensity of the optical power.

In order to observe the $\mathcal{PT}$-symmetric and broken $\mathcal{PT}$-symmetric behaviours, besides changing the optically induced gain (or damping) by tuning the amplitude of the driving field as shown above, we can also change the coupling between the two mechanical resonator from weak to strong by applying stress to the coupling overhang by piezoelectric effect or photothermal effect~\cite{KarabalinPRB09,OkamotoAPE09,OkamotoPE10,OkamotoAPL11,OkamotoNaP13}, as shown in Fig.~\ref{fig2}.

\section{mechanical $\mathcal{PT}$-symmetry in quantum regime}

Up to now the dynamical behavior of the system is obtained by treating the operators as c-numbers in the semi-classical approximation. This theory is only applied in the condition that the numbers of photons and phonons in the system are so large that the fluctuations of the cavity fields and mechanical resonators can be neglected.
With the progress in the experiments, the mechanical resonators have already been cooled and prepared near the ground states in the optomechanical systems~\cite{Teufel475,Chan2009}, then the thermal excitations and even the quantum fluctuations should be considered in the derivation. In this section, we are going to develop a description of mechanical $\mathcal{PT}$-symmetric systems by using full quantum theory~\cite{AgarwalPRA13}.

The linearized quantum Langevin equations (LQLEs) for the operators [with adding quantum noises to the Eqs.~(\ref{eq:15})-(\ref{eq:20})] are given as
\begin{equation}   \label{eq:a33}
\frac{d}{dt}V=MV+F,
\end{equation}
where $V=\left(
\begin{array}{cccccccc}
a_{1},a_{2},a_{1}^{\dag },a_{2}^{\dag },q_{1},q_{2},p_{1},p_{2}%
\end{array}%
\right) ^{T}$, $F=\left(
\begin{array}{cccccccc}
\sqrt{\kappa _{1}}a_{1,in},\sqrt{\kappa _{2}}a_{2,in},\sqrt{\kappa _{1}}a_{1,in}^{\dag },\sqrt{\kappa _{2}}a_{2,in}^{\dag },0,0,\xi _{1},\xi _{2}%
\end{array}%
\right) ^{T}$, and
\begin{widetext}
\begin{equation} \label{eq:a34}
M=\left(
\begin{array}{cccccccc}
-\left( \frac{\kappa _{1}}{2}+i\Delta _{1}^{\prime }\right)  & 0 & 0 & 0 &
iG_{1} & 0 & 0 & 0 \\
0 & -\left( \frac{\kappa _{2}}{2}+i\Delta _{2}^{\prime }\right)  & 0 & 0 & 0
& iG_{2} & 0 & 0 \\
0 & 0 & -\left( \frac{\kappa _{1}}{2}-i\Delta _{1}^{\prime }\right)  & 0 &
-iG_{1}^{\ast } & 0 & 0 & 0 \\
0 & 0 & 0 & -\left( \frac{\kappa _{2}}{2}-i\Delta _{2}^{\prime }\right)  & 0
& -iG_{2}^{\ast } & 0 & 0 \\
0 & 0 & 0 & 0 & 0 & 0 & \omega _{1} & 0 \\
0 & 0 & 0 & 0 & 0 & 0 & 0 & \omega _{2} \\
G_{1}^{\ast } & 0 & G_{1} & 0 & -\omega _{1} & J & -\frac{\gamma _{1}}{2} & 0
\\
0 & G_{2}^{\ast } & 0 & G_{2} & J & -\omega _{2} & 0 & -\frac{\gamma _{2}}{2}%
\end{array}%
\right) .
\end{equation}
\end{widetext}
The quantum noise $a_{i,in}$ $\left( i=1,2\right) $ satisfies
the communication relation $\left[ a_{i,in}\left( t\right) ,a_{i,in}^{\dag }\left( t^{\prime }\right) \right] =\delta \left( t-t^{\prime }\right) $ and
the correlations $\left\langle a_{i,in}^{\dag }\left( t\right)
a_{i,in}\left( t^{\prime }\right) \right\rangle =0,$ and $\left\langle
a_{i,in}\left( t\right) a_{i,in}^{\dag }\left( t^{\prime }\right)
\right\rangle =\delta \left( t-t^{\prime }\right) $; the Brownian stochastic
force $\xi _{i}$ $\left( i=1,2\right) $ with zero mean value satisfies the correlation $\left\langle \xi _{i}\left( t\right) \xi _{i}\left( t^{\prime }\right)
\right\rangle =\gamma _{i}\left( n^{\rm{th}}_{i}+1/2\right) \delta \left(
t-t^{\prime }\right) $ with the thermal phonon number $n^{\rm{th}}_{i}=\left( {\rm exp} \left\{ \hbar \omega _{i}/k_{B}T\right\} -1\right) ^{-1}$,
where $k_{B}$ is the Boltzmann constant and $T$ is the effective temperature of the reservoir of the mechanical oscillators.

\begin{figure}[tbp]
\includegraphics[bb=30 332 557 535, width=8.5 cm, clip]{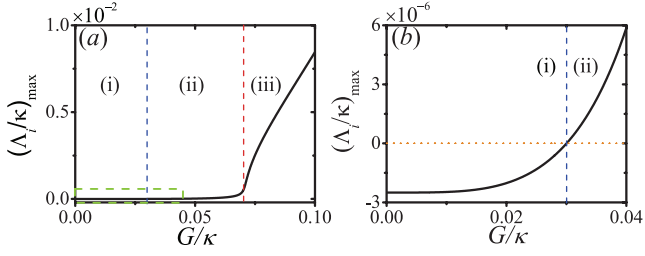}
\caption{(Color online) The maximum value $\left( \Lambda_{i}/\kappa\right) _{\rm max }$ of all the eigenvalues of $M$ as a function of the effective optomechanical coupling rate $G_1=G_2 \equiv G$. The parameters are $\kappa _{1}=\kappa _{2}=\kappa$, $J=\kappa/100$, $-\Delta_{1}^{\prime }=\Delta_{2}^{\prime }=\omega _{1}=\omega _{2}=10\kappa$ and $\gamma _{1}=\gamma _{2} =\kappa/10^{5}$.}
\label{fig8}
\end{figure}

The system is stable only if the real parts of all the eigenvalues $\Lambda_{i}$ ($i=1,2,\ldots,8$) of matrix $M$ are negative ($\Lambda_{i}<0$). The stability conditions can be given explicitly by using the Routh-Hurwitz criterion~\cite{DeJesusPRA87}, but they are too cumbersome to be given here. In the following, we will analyze the stability conditions of the system numerically. The maximum value $\left( \Lambda_{i}\right) _{\rm max }$ of all the eigenvalues of $M$ as a function of the effective optomechanical coupling rate $G_1=G_2 \equiv G$ is shown in Fig.~\ref{fig8}. For the parameters under consideration in this figure, there are three regions: (i) $0\leq G/\kappa \lesssim 0.03$: $\left( \Lambda _{i}\right) _{\rm max }<0$, the system is stable in this region (stable region); (ii) $0.03\lesssim G/\kappa\lesssim 0.07 $: $\left( \Lambda _{i}\right)_{\rm max }$ ($>0$) increases slowly with the effective optomechanical coupling rate $G$, and the system can be stable in a long time because $\left( \Lambda _{i}\right)_{\rm max }$ is small in the region (we can call this region quasi-stable region); (iii) $0.07 \lesssim G/\kappa$, $\left( \Lambda _{i}\right) _{\rm max }$ ($>0$) increases fast with the effective optomechanical coupling rate $G$, i.e., this is the unstable region. It is worth mentioning that $G/\kappa \approx 0.07 $ is also the exceptional point at which $\gamma_{\rm eff} \approx 2J$ and the mechanical $\mathcal{PT}$-symmetry is broken.

The solution to the LQLEs (\ref{eq:a33}) is given by~\cite{AgarwalPRA13}
\begin{equation}  \label{eq:a35}
V\left( t\right) =K\left( t\right) V\left( 0\right)
+\int_{0}^{t}K\left(t-t^{\prime }\right) F\left( t^{\prime }\right) dt^{\prime }
\end{equation}%
with $K\left( t\right) =e^{Mt}$. The total phonons generated in each mechanical resonators $n^{\rm tt}_{i}(t)=\left[\left\langle q_{i}^{2}(t)\right\rangle+ \left\langle p_{i}^{2}(t)\right\rangle -1\right]/2$ come from two parts,
\begin{equation}  \label{eq:a36}
n^{\rm tt}_{i}(t)=n^{\rm st}_{i}(t)+n^{\rm sp}_{i}(t),
\end{equation}
where $i=1,2$,
\begin{widetext}
\begin{eqnarray}
n^{\rm st}_{1}(t) &=&\frac{1}{2}\sum_{j=1,3}\sum_{i=1,2}\left\{ \left\vert
K_{4+j,i}(t)\right\vert ^{2}\left[ 2\left\langle a_{i}^{\dagger
}(0)a_{i}(0)\right\rangle +1\right] +\left\vert K_{4+j,4+i}(t)\right\vert
^{2}\left\langle q_{i}^{2}(0)\right\rangle +\left\vert
K_{4+j,6+i}(t)\right\vert ^{2}\left\langle p_{i}^{2}(0)\right\rangle
\right\} -\frac{1}{2}, \label{eq:a37}\\
n^{\rm st}_{2}(t) &=&\frac{1}{2}\sum_{j=2,4}\sum_{i=1,2}\left\{ \left\vert
K_{4+j,i}(t)\right\vert ^{2}\left[ 2\left\langle a_{i}^{\dagger
}(0)a_{i}(0)\right\rangle +1\right] +\left\vert K_{4+j,4+i}(t)\right\vert
^{2}\left\langle q_{i}^{2}(0)\right\rangle +\left\vert
K_{4+j,6+i}(t)\right\vert ^{2}\left\langle p_{i}^{2}(0)\right\rangle
\right\} -\frac{1}{2}, \label{eq:a38}\\
n^{\rm sp}_{1}(t) &=&\frac{1}{2}\sum_{j=1,3}\sum_{i=1,2}\left\{ \kappa
_{i}\int_{0}^{t}\left\vert K_{4+j,2+i}(t^{\prime })\right\vert
^{2}dt^{\prime }+\gamma _{i}\left( n_{i,\mathrm{th}}+1/2\right)
\int_{0}^{t}\left\vert K_{4+j,6+i}(t^{\prime })\right\vert ^{2}dt^{\prime
}\right\},  \label{eq:a39}\\
n^{\rm sp}_{2}(t) &=&\frac{1}{2}\sum_{j=2,4}\sum_{i=1,2}\left\{ \kappa
_{i}\int_{0}^{t}\left\vert K_{4+j,2+i}(t^{\prime })\right\vert
^{2}dt^{\prime }+\gamma _{i}\left( n_{i,\mathrm{th}}+1/2\right)
\int_{0}^{t}\left\vert K_{4+j,6+i}(t^{\prime })\right\vert ^{2}dt^{\prime
}\right\} \label{eq:a40}
\end{eqnarray}
with $K_{j,i}\left( t\right) =\left(e^{Mt}\right)_{j,i}$.

\begin{figure*}[tbp]
\includegraphics[bb=20 249 571 687, width=13 cm, clip]{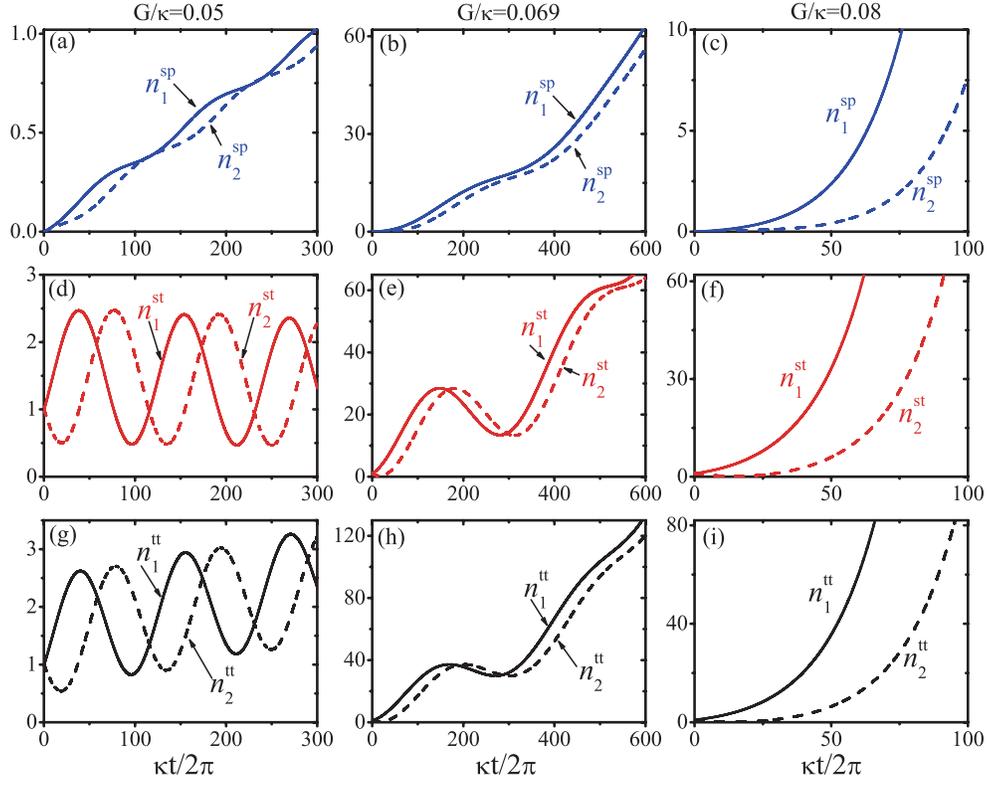}
\caption{(Color online) The dynamical behaviours of the phonons by spontaneous generation $n^{\rm sp}_{1}$ (blue solid line) and $n^{\rm sp}_{2}$ (blue dash line) [(a), (b) and (c)], stimulated generation $n^{\rm st}_{1}$ (red solid line) and $n^{\rm st}_{2}$ (red dash line) [(d), (e) and (f)] and the total phonons $n^{\rm tt}_{1}$ (black solid line) and $n^{\rm tt}_{2}$ (black dash line) [(g), (h) and (i)] with different values of the effective optomechanical coupling rate $G_1=G_2 \equiv G$: (a), (d) and (g) $G/\kappa=0.05$; (b), (e) and (h) $G/\kappa=0.069$; (c), (f) and (i) $G/\kappa=0.08$. Here $\langle q^{2}_{i}(0) \rangle=\langle p^{2}_{i}(0)\rangle  =3/2$, $T=0$, i.e., $n^{\rm{th}}_{i}=0$. The other parameters are the same as that given in Fig.~\ref{fig8}.}
\label{fig9}
\end{figure*}
\end{widetext}

From Eqs.~(\ref{eq:a35}) and (\ref{eq:a36}), the total phonons $n_{i}^{\rm tt}$ ($i=1,2$) include contributions from both the stimulated and spontaneous generations, where the stimulated generation $n_{i}^{\rm st}$ comes from the term $K\left( t\right) V\left( 0\right)$ and the spontaneous generation $n_{i}^{\rm sp}$ is contributed by the term $\int_{0}^{t}K\left(t-t^{\prime }\right) F\left( t^{\prime }\right) dt^{\prime }$. The number of the phonons by the spontaneous generation is shown in Fig.~\ref{fig9} (a), (b) and (c); the one by the stimulated generation is shown in Fig.~\ref{fig9} (d), (e) and (f) with the initial condition that there is no photon in the optical modes and one phonon in each mechanical resonators initially, i.e., $\langle a^{\dagger}_{i}(0)a_{i}(0) \rangle =0$ and $\langle q^{2}_{i}(0) \rangle=\langle p^{2}_{i}(0)\rangle  =3/2$ ($i=1,2$). The total phonons generated in the mechanical resonators are shown in Figs.~\ref{fig9} (g), (h) and (i).

From Figs.~\ref{fig9} (a), (d) and (g), as the optomechanical coupling rate $G$ is in the $\mathcal{PT}$-symmetry region ($G/\kappa=0.05$), the number of the phonons by the spontaneous generation increases monotonously (in a series of cascades) and the number of the phonons by the stimulated generation shows some oscillation behavior. The phonons by the spontaneous generation will dominate the total generation of phonons after an enough long time. When the optomechanical coupling rate $G$ is in the unstable region ($\mathcal{PT}$-symmetry broken region) as shown in Figs.~\ref{fig9} (c), (f) and (i), the number of the phonons by both the stimulated and spontaneous generations increases exponentially. The phonons generated by the spontaneous generation still play an important role in the total generation of the phonons. So the effect of the quantum noises can not be ignored when the number of the phonons by stimulated generation is small and the evolution time is long enough.

\begin{figure}[tbp]
\includegraphics[bb=56 230 513 585, width=8.5 cm, clip]{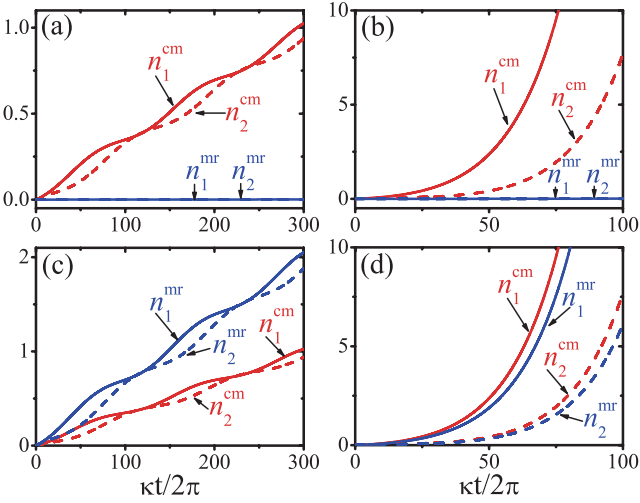}
\caption{(Color online) The dynamical behaviours of the phonons generated by the quantum noises of the cavity modes $n^{\rm cm}_{i}$ and by the stochastic forces of the mechanical resonators $n^{\rm mr}_{i}$ with thermal phonon number: (a) and (b) $n^{\rm{th}}_{i}=0$; (c) and (d) $n^{\rm{th}}_{i}=1000$. The effective optomechanical coupling rate $G_1=G_2 \equiv G$ is: (a) and (c) $G/\kappa=0.05$; (b) and (d) $G/\kappa=0.08$. The other parameters are the same as that given in Fig.~\ref{fig8}.}
\label{fig10}
\end{figure}

Moreover, the phonons by the spontaneous generation originate from two sources as shown in Fig.~\ref{fig10} or Eqs.~(\ref{eq:a39}) and (\ref{eq:a40}): the first term results from the quantum noises of the cavity modes (labeled as '$n_{i}^{\rm cm}$') and the stochastic forces of the mechanical resonators contribute the second term (labeled as '$n_{i}^{\rm mr}$'). Under zero temperature as shown in Fig.~\ref{fig10} (a) and (b), since $\gamma_{1}=\gamma_{2} \ll \kappa_{1}=\kappa_{2}$, the phonons by the spontaneous generation mainly come from the quantum noises of the cavity modes. When the thermal phonon number $n^{\rm{th}}_{i}$ approaches about $1000$ as shown in Figs.~\ref{fig10} (c) and (d), the contributions by the stochastic force of the mechanical resonators can be comparable with the one by the quantum noises of the cavity modes.

\begin{figure}[tbp]
\includegraphics[bb=93 273 512 598, width=8.5 cm, clip]{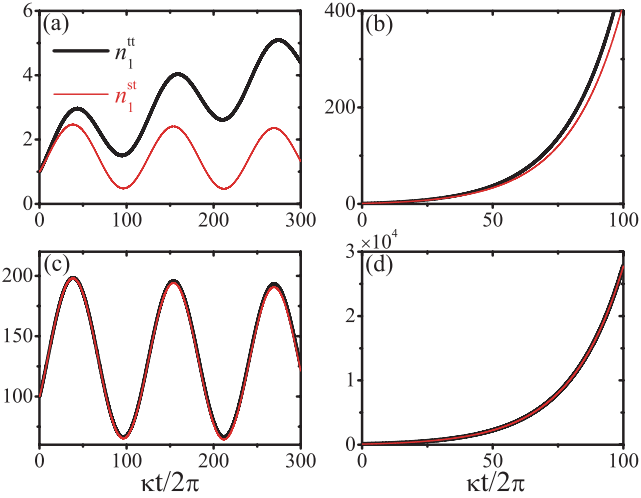}
\caption{(Color online) The dynamical behaviours of the phonons generated by the stimulated generation $n^{\rm st}_{1}$ (red thin line) and the total phonons $n^{\rm tt}_{1}$ (black bold line) with thermal phonon number $n^{\rm{th}}_{i}=1000$: (a) and (b) $\langle q^{2}_{i}(0) \rangle=\langle p^{2}_{i}(0)\rangle  =3/2$; (c) and (d) $\langle q^{2}_{i}(0) \rangle=\langle p^{2}_{i}(0)\rangle  =100\frac{1}{2}$. The effective optomechanical coupling rate is $G/\kappa=0.05$ in (a) and (c), and is $G/\kappa=0.08$ in (b) and (d). The other parameters are the same as that given in Fig.~\ref{fig8}.}
\label{fig11}
\end{figure}

Lastly we will numerically reveal that the phonons by the spontaneous generation can be neglected when the number of phonons are large initially. The dynamical behaviours of the phonons by the stimulated generation $n^{\rm st}_{1}$ and the total phonons $n^{\rm tt}_{1}$ in mechanical resonator $1$ are as shown in Fig.~\ref{fig11}, where thermal phonon number is $n^{\rm{th}}_{i}=1000$.
Unlike in the condition that the number of phonons is small initially [Fig.~\ref{fig11} (a) and (b)], where the phonons generated by the spontaneous generation play an important role in the total generation of the phonons, the phonons generated by the spontaneous generation can be neglected when the the number of phonons in the mechanical resonator approaches about $100$ initially as shown in Fig.~\ref{fig11} (c) and (d).

\section{discussions and conclusions}

Let us now discuss the experimental feasibility for the observation
of mechanical $\mathcal{PT}$-symmetry in the coupled optomechanical
systems. In the calculation, we have assumed that the parameters satisfy the following conditions: (i) resolved-sideband condition, $\omega_{m}=10 \kappa$; (ii) weak optomechanical coupling, $g=\kappa/10000$, $G_{i}=\kappa/20$; (iii)
strong driving condition, $\Omega=5000\kappa$ to $10000\kappa$; (iv) resonant blue- and red-sideband conditions, $\Delta _{1}=-\omega _{m}$, $\Delta _{2}=\omega _{m}$; (v) high mechanical quality factor such that $\gamma _{i} \ll \Gamma _{i}$. Most of the parameters used in the calculation are within the reach of the current technology. For example, the sideband cooling of mechanical resonator has been observed in many different types of optomechanical
systems~\cite{Schliesser2008,Rocheleau, Groblacher2009, Park,
Schliesser2009, Teufel475,Chan2009, Verhagen}; the optomechanical coupling
constant has been reported to reach the level of $g=10^{-2}\kappa$ in the zipper
cavity and double-disk cavity~\cite{PainterNat09,LinPRL09,LinNPh10,WiederheckerNat09}.

In summary, we have theoretically demonstrated that the coupled optomechanical systems can be used to observe the $\mathcal{PT}$ symmetry for the mechanical degrees of freedom. The dynamical equations for two coupled mechanical resonators with $\mathcal{PT}$ symmetry are derived by adiabatically eliminating the degrees of freedom of the cavity modes. By tuning the amplitudes of the driving fields or the coupling constant between the vibrating mirrors, we can observe the transition between the $\mathcal{PT}$-symmetric and broken $\mathcal{PT}$-symmetric phases. In the $\mathcal{PT}$-symmetric region, the two vibrating mirrors become two beat frequency oscillators. While in the broken $\mathcal{PT}$-symmetric region, the oscillation amplitudes of the mirrors increase (or after an initial decrease) exponentially, which might result in the photon lasing. In the experiment, mechanical $\mathcal{PT}$-symmetric phase transition can be demonstrated by measuring the evolution of the output intensity of the cavities.

Additionally, when the number of phonons is small, we consider the contributions of the quantum noises by the LQLEs~\cite{AgarwalPRA13,HePRA15}. In the $\mathcal{PT}$-symmetric region, the phonons by spontaneous generation dominate the total phonon generation after an enough long time; when the $\mathcal{PT}$-symmetric is broken, the phonons by spontaneous generation still remain important in the total phonon generation. The coupled optomechanical systems offer us a potential platform to push mechanical resonators into the quantum regime and our proposal may pave the way to study the mechanical $\mathcal{PT}$-symmetric systems in the quantum regime~\cite{AgarwalPRA13,DastPRA14,HePRA15}.

\section{Acknowledgement}

X.W.X. thanks W. Tan, K. Li, W. J. Nie, Q. Zheng, L. Ge and Y. Yao for helpful discussions and comments. This work is supported by the Postdoctoral Science Foundation of China (under Grant No. 2014M550019), the National Natural Science Foundation of China (under Grants No. 11422437, No. 11174027, and No. 11121403) and the 973 program (under Grants No. 2012CB922104 and No. 2014CB921403). Y.X.L. is supported by the National Natural Science Foundation of China (under Grant Nos. 61025022, 61328502).

\appendix

\section{Adiabatical elimination}

From Eqs.~(\ref{eq:15})-(\ref{eq:20}), we can derive the dynamical equations
for the coupled mechanical resonators by adiabatically eliminating the
cavity modes. As $q_{i} = ( b_{i}^{\dag }+b_{i} )/\sqrt{2}$, Eq.~(\ref{eq:15}%
) can be rewritten as
\begin{eqnarray}
\frac{d}{dt}a_{i}&=&-\left( \frac{\kappa _{i}}{2}+i\Delta_{i}^{\prime
}\right) a_{i}+i\frac{G _{i}}{\sqrt{2}}\left( b_{i}^{\dag }+b_{i} \right).
\label{eq:22}
\end{eqnarray}
In order to observe the $\mathcal{PT}$ symmetry in the coupled vibrating mirrors,
we need to provide gain to the left vibrating mirror and equivalent damping to
the right vibrating mirror respectively. Therefore, we assume that the driving
field to the left cavity is resonant to the blue sideband ($\Delta
_{1}^{\prime }=-\omega _{1}$) and the driving field to the right cavity is
resonant to the red sideband ($\Delta _{2}^{\prime }=\omega _{2}$). After
introducing the slowly varying amplitudes: $\widetilde{a}_{1}
=a_{1}e^{\left( \frac{\kappa _{1}}{2}-i\omega_{1}\right) t} $, $\widetilde{a}%
_{2} =a_{2}e^{\left( \frac{\kappa _{2}}{2}+i\omega_{2}\right) t} $, $%
\widetilde{b}_{1} =b_{1}e^{i\omega _{1}t} $, $\widetilde{b}_{2}
=b_{2}e^{i\omega _{2}t} $, and integrating Eq.~(\ref{eq:22}), we get
\begin{eqnarray}
\widetilde{a}_{1} &=&\int_{-\infty }^{t}i\frac{G_{1}}{\sqrt{2}}\left(%
\widetilde{b}_{1}^{\dag }+\widetilde{b}_{1}e^{-i2\omega _{1}\tau }\right) e^{%
\frac{\kappa _{1}}{2}\tau }d\tau , \\
\widetilde{a}_{2} &=&\int_{-\infty }^{t}i\frac{G_{2}}{\sqrt{2}}\left(%
\widetilde{b}_{2}+\widetilde{b}_{2}^{\dag }e^{i2\omega _{2}\tau }\right) e^{%
\frac{\kappa _{2}}{2}\tau }d\tau .
\end{eqnarray}%
Under the assumption that the decay rates of the cavities are much larger
than the effective optomechanical coupling $\kappa _{i} \gg G _{i}$, we can
adiabatically eliminate the cavity modes. As $\kappa _{i} \gg \gamma _{i}$, the evolution of $
\widetilde{b}_{i}$ is much slower than $\widetilde{a}_{i}$, so that we
can take $\widetilde{b}_{i}$ out of the integrals and evaluate the integrals
directly, then we find the approximate expressions of the cavity modes~\cite%
{JahnePRA09},
\begin{eqnarray}
a_{1} & = &i\frac{\sqrt{2}G_{1}}{\kappa _{1}}b_{1}^{\dag }+i\frac{\sqrt{2}%
G_{1}}{\left( \kappa _{1}-i4\omega _{1}\right) }b_{1}, \\
a_{2} & = &i\frac{\sqrt{2}G_{2}}{\kappa _{2}}b_{2}+i\frac{\sqrt{2}G_{2}}{%
\left( \kappa _{2}+i4\omega _{2}\right) }b_{2}^{\dag }.
\end{eqnarray}%
Substituting these expressions into Eqs.~(\ref{eq:19}) and (\ref{eq:20}),
then we find Eqs.~(\ref{eq:27}) and (\ref{eq:28}).

\section{Bi-orthogonal basis approach}

We are going to solve Eqs.~(\ref{eq:35}) analytically by the bi-orthogonal
basis approach~\cite{WongJMP67,SunPS93,LeungPRE98}. In order to write these
solutions in an explicit form, we use the engenstates
\begin{equation}
\left\vert 1\right\rangle =\left(
\begin{array}{c}
1 \\
0 \\
0 \\
0%
\end{array}%
\right) , \left\vert 2\right\rangle =\left(
\begin{array}{c}
0 \\
1 \\
0 \\
0%
\end{array}%
\right) , \left\vert 3\right\rangle =\left(
\begin{array}{c}
0 \\
0 \\
1 \\
0%
\end{array}%
\right) , \left\vert 4\right\rangle =\left(
\begin{array}{c}
0 \\
0 \\
0 \\
1%
\end{array}%
\right).
\end{equation}%
The effective Hamiltonian $H_{\mathrm{eff}}$ satisfy the eigenvalue
equations
\begin{eqnarray}
H_{\mathrm{eff}}\left\vert \phi _{i}\right\rangle &=&\lambda _{i}\left\vert
\phi _{i}\right\rangle, \\
H_{\mathrm{eff}}^{\dag }\left\vert \varphi _{i}\right\rangle &=&\lambda
_{i}^{\ast }\left\vert \varphi _{i}\right\rangle,
\end{eqnarray}
where $\left\vert \phi _{i}\right\rangle$ and $\left\vert \varphi
_{i}\right\rangle$ are called the bi-orthogonal basis, and they satisfy the
bi-orthogonal relations
\begin{equation}
\left\langle\varphi _{i} \vert \phi _{j}\right\rangle =D_{i}\delta _{ij},
\end{equation}
and generalized completeness relations
\begin{equation}
\sum_{i}\frac{\left\vert \phi _{i}\right\rangle \left\langle \varphi
_{i}\right\vert }{\left\langle \varphi _{i} \vert \phi _{i}\right\rangle }%
=\sum_{i}\frac{\left\vert \varphi _{i}\right\rangle \left\langle \phi
_{i}\right\vert }{\left\langle \phi _{i} \vert \varphi _{i}\right\rangle }=1.
\end{equation}
So we have the basis transform relations
\begin{equation}
\left\vert n\right\rangle=\sum_{i}\frac{ \left\langle \varphi _{i} \vert
n\right\rangle }{\left\langle \varphi _{i} \vert \phi _{i}\right\rangle }%
\left\vert \phi _{i}\right\rangle =\sum_{i}\frac{ \left\langle \phi _{i}
\vert n\right\rangle }{\left\langle \phi _{i} \vert \varphi
_{i}\right\rangle }\left\vert \varphi _{i}\right\rangle,
\end{equation}
with
\begin{equation}
X_{n,i}=\left\langle n \vert \phi _{i}\right\rangle, Y_{n,i}=\left\langle n
\vert \varphi _{i}\right\rangle,
\end{equation}%
and
\begin{equation}  \label{eq:46}
X_{i}=\left[ \frac{J}{\lambda _{i}}\chi_{i}, -i\frac{J}{\omega _{m}}%
\chi_{i}, i\frac{\omega _{m}}{\lambda _{i}}, 1\right]^{T},
\end{equation}%
\begin{equation}
Y_{i}=\left[ -i\frac{J}{\lambda _{i}^{\ast }} -\frac{J}{\lambda _{i}^{\ast}}%
\chi_{i}^{\ast}, i\frac{J}{\omega _{m}}\chi_{i}^{\ast}, i\frac{\omega _{m}}{%
\lambda _{i}^{\ast}} +\frac{J^{2}}{\omega _{m}\lambda _{i}^{\ast}}%
\chi_{i}^{\ast}, 1 \right]^{T},
\end{equation}
where
\begin{equation}
\chi_{i} = \frac{\omega _{m}^{2}}{\frac{1}{2}\gamma _{\mathrm{eff}}\lambda
_{i}+i\lambda_{i}^{2}-i\omega _{m}^{2}}.
\end{equation}

In terms of the biorthonormal basis \{$\left\vert \phi _{i}\right\rangle$, $%
\left\vert \varphi _{i}\right\rangle$\}, the initial state is given as
\begin{eqnarray}
\left\vert \Psi \left( 0\right) \right\rangle &=&\sum_{n}c_{n}\left\vert
n\right\rangle  \notag \\
&=&\sum_{i}\sum_{n}\frac{c_{n}\left\langle \varphi _{i} \vert n\right\rangle
}{\left\langle \varphi _{i} \vert \phi _{i}\right\rangle }\left\vert \phi
_{i}\right\rangle
\end{eqnarray}%
where the initial conditions are: $c_{1}=q_{1}$, $c_{2}=p_{1}$, $c_{3}=q_{2}$
and $c_{4}=p_{2}$. Then we have
\begin{eqnarray}
\left\vert \Psi \left( t\right) \right\rangle &=&e^{-iH_{\mathrm{eff}%
}t}\left\vert \Psi \left( 0\right) \right\rangle  \notag \\
&=&\sum_{i}\sum_{n}\frac{c_{n}\left\langle \varphi _{i} \vert n\right\rangle
}{\left\langle \varphi _{i} \vert \phi _{i}\right\rangle }e^{-i\lambda
_{i}t}\left\vert \phi _{i}\right\rangle  \notag \\
&=&\sum_{k}\sum_{n}\sum_{i}c_{n}\frac{\left\langle k \vert \phi
_{i}\right\rangle e^{-i\lambda _{i}t}\left\langle \varphi _{i} \vert
n\right\rangle }{\left\langle \varphi _{i} \vert \phi _{i}\right\rangle }%
\left\vert k\right\rangle.
\end{eqnarray}
The dynamics of the oscillators are given as
\begin{eqnarray}
q_{1}(t)=\sum_{n}\sum_{i}c_{n}\frac{\left\langle 1 \vert \phi
_{i}\right\rangle e^{-i\lambda _{i}t}\left\langle \varphi _{i} \vert
n\right\rangle }{ \left\langle \varphi _{i} \vert \phi _{i}\right\rangle },
\\
q_{2}(t)=\sum_{n}\sum_{i}c_{n}\frac{\left\langle 3 \vert \phi
_{i}\right\rangle e^{-i\lambda _{i}t}\left\langle \varphi _{i} \vert
n\right\rangle }{ \left\langle \varphi _{i} \vert \phi _{i}\right\rangle }.
\end{eqnarray}

\bibliographystyle{apsrev}
\bibliography{ref}

\end{document}